# Feasibility of sensor-based technology for monitoring health in developing countries - cost analysis and user perception aspects

*Adelina Basholli[1], Thomas Lagkas[2], Peter A. Bath[3], and George Eleftherakis[2]*

[1]South-East European Research Centre, University of Sheffield, Thessaloniki, Greece
[2]CITY College- International Faculty of the University of Sheffield, Thessaloniki, Greece
[3]Information School, University of Sheffield, Sheffield, U.K.
adbasholli@seerc.org, {tlagkas, eleftherakis}@city.academic.gr, p.a.bath@sheffield.ac.uk

*Understanding the financial burden of chronic diseases in developing regions still remains an important economical factor which influences the successful implementation of sensor based applications for continuous monitoring of chronic conditions. Our research focused on a comparison of literature-based data with real costs of the management and treatment of chronic diseases in a developing country, and we are using Kosovo as an example here. The results reveal that the actual living costs exceed the minimum expenses that chronic diseases impose. Following the potential of a positive economic impact of sensor based platforms for monitoring chronic conditions, we further examined the users' perception of digital technology. The purpose of this paper is to present the varying cost levels of treating chronic diseases, identify the users' concerns and requirements towards digital technology and discuss issues and challenges that the application of sensor based platforms imply in low- and middle-income countries.*

## 1. Introduction

The basic medical care and continuous monitoring of health conditions present a need that is in focus especially nowadays. The globalizations of economies, the tendency of people to live in cities, and the evolved technology trends, have caused real impact in the health system. Especially regarding developing regions, it was concluded [1, 2] that many elder people are living with a possible chronic disease, such as heart disease, diabetes, and Alzheimer's disease. Moreover, chronic conditions constitute one of the leading factors that are causing nearly 80% of deaths in developing countries [3], affecting not only the elderly population but the working age people, too [2].

The percentage of individuals suffering from chronic diseases is expected to rise up to 65% of general population by the year 2030, especially in low- and middle-income countries [2]. Furthermore, the public system in these regions can be particularly affected by the global spread of chronic diseases, and thus leading to a high possibility of having to deal with a number of negative consequences that chronic conditions cause. Towards helping developing regions cope with the increased needs for treatment of patients with chronic diseases, digital technology can be considered as a possible solution.

E-health presents the discipline of applying Information and Communication Technology (ICT) in health services. Under the umbrella of the term e-health the following categories of services can be included: support of healthcare facilities, ability of continuous monitoring, and health education and knowledge. Sensor based networks are worth investigating as an application that can potentially be applied in all aforementioned categories. Application of sensor based networks can enhance the provision of healthcare in developing countries, improve health delivery and facilitate the connection of patients, whose living conditions do not allow them immediate access to health services, thus increasing the number of positive patient outcomes [4, 5]. Various studies consider health information technology as a positive method applied in health system (around 92 percent of related articles [6]); and suggest that technology in healthcare has already emerged, with the increasing number of applications that are being developed continuously. Moreover, Mishra et al. [5] concluded that the implementation of digital technology in healthcare results in effectively reducing cost of treatment compared to traditional methods of treatment, especially for developing countries.

Continuous and at-home health services will provide better life to patients diagnosed with chronic diseases, and address their social and financial loads [1]. Sensor-based networks are considered as a possible solution that may provide a better and more affordable method applied in healthcare [2]. These technologies have many advantages compared to traditional health offering tools, such as: ease of use, lower possibility of failure, reduced risk of infection, user comfortability, enhanced mobility, and decreased cost of delivery. Furthermore, their spread and usage is increasing and by year 2019 their market share is expected to reach 14.6 billion dollars [7].

World-wide the number of aging population is increasing year-by-year. This is also the case with developing countries like Kosovo [8]. Furthermore, Kosovo health system is facing a high number of patients affected by chronic conditions while having very limited space and resources to treat them [9]. This is shown also in the below figure.

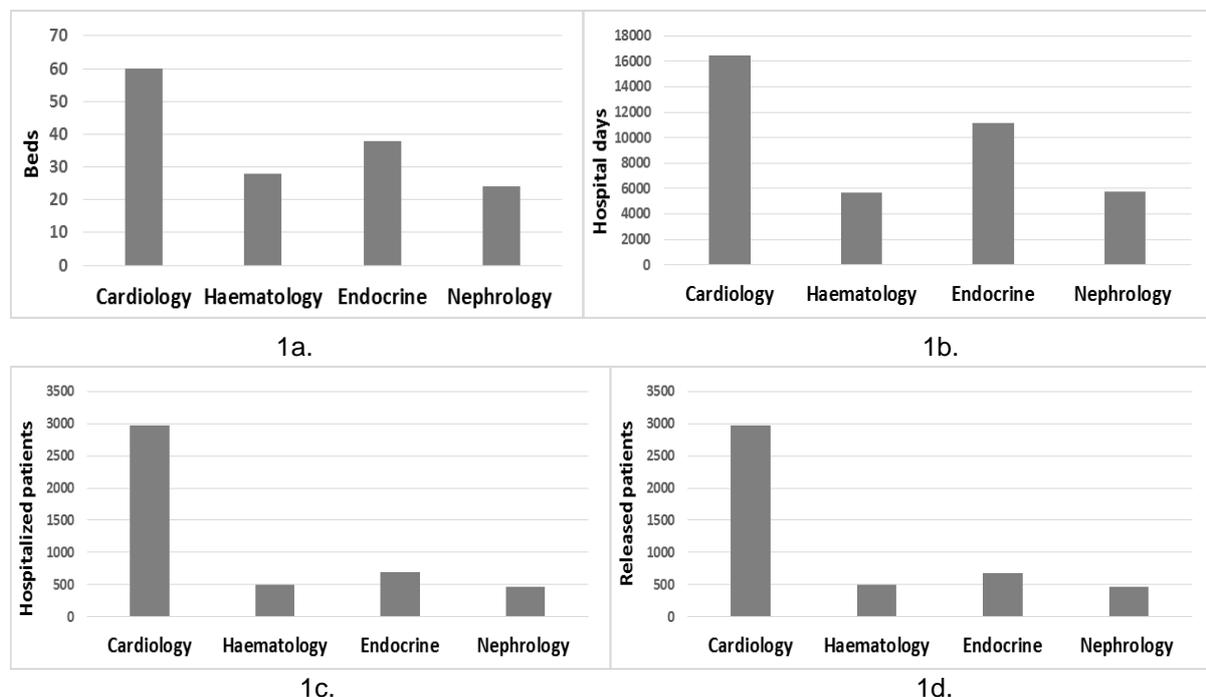

**Figure 1** Comparison of patient's requests for health services with available resources in Kosovo clinics [9]

Figure 1 presents the difference in the number of beds compared with the total number of days spend in hospital. For example, comparing the number of available beds (1a) in the clinic of Cardiology with the number of hospitalized patients (1c) or days spend in the hospital (1b), we can notice around 3000 patient's requests which can be hospitalized in just 60 available beds. In this figure are included just some of the clinics in Kosovo, where various chronic conditions are treated, such as: Cardiology, Haematology, Endocrine, and Nephrology Clinic. Besides the limited space and resources to treat chronic ill patients, Kosovo health system has also limited number of medical staff. From the same report [9] it is shown that the total number of doctors from the four considered clinics is only 44.

Regardless of its health system, Kosovo's network infrastructure and offered services are developing day-by-day. This is evident also by the high percentage of internet penetration (76.6%) which is comparable with the developed countries [10]. More details about the Kosovo health system and its network architecture can be found in [11].

The purpose of this paper is twofold. Our first aim is to present a comparison related with the financial impact of using sensor based platforms in developing countries using Kosovo as an example. The gathered data from similar studies are compared with the average person income and corresponding health expenses for treating chronic diseases. Secondly, seeing the benefits and potential profits of applying sensor-based platforms in a low- and middle-income country, we review literature on the users' acceptance of digital technology in order to identify the key factors and apply a similar investigation in Kosovo too. To the best of our knowledge, there are just a few studies in the literature that evaluate the perception of users while specifically applying sensor-based applications. This research will assist us in our future work to investigate the patient and healthcare staff perception of sensor-based applications for monitoring chronic diseases in developing countries. Furthermore we plan to elaborate user's experience with possible sensor-based technologies, their perception of applying technology in health sector and possible implications of this technology based on patient's categories (age, gender, disease heaviness, economic situation, etc).

The outline of this paper is as follows: Section 2 presents a thorough cost analysis, together with economic consequences of chronic diseases, possible solutions that will lower the cost for treating them and corresponding information about the living costs and health expenses in Kosovo. Section 3 investigates existing literature on the user perception of digital technology from the patients' and health professionals' point of view. Section 4 discusses possible issues and challenges needed to be addressed for future studies. Finally, section 5 provides the overall conclusions of this paper.

## 2. Cost Analysis

Every year around 9 million people under 60 years old die due to chronic diseases, and 90% of these deaths occur in low- and middle-income countries [3]. However, not only the developing countries are dealing with the struggle of managing and treating chronic conditions. Developed countries also spend high amounts of money to handle chronic diseases. For example, the United States of America spends around 2 trillion dollars per year [1]. The cost of heart disease or stroke was reported to be around 394 billion dollars in 2008 [1]. This comparison with a developed country, like United States of America, is presented just to show the economic burden that developing countries will have it even harder to cope with.

Considering the increased number of aging population, where the older population is expected to outnumber the younger one, the cost for treating chronic diseases are expected to become even higher. Therefore, Suhrcke et al. [2] estimated the costs of chronic disease measured as cost-of-illness and linked them to the countries' Gross Domestic Product (GDP). Hence, for developed countries the cost of chronic disease varies from 1-3% of GDP, whereas for the developing countries the values vary between 1.8% for Venezuela and 5.9% for Barbados.

We should also consider that chronic diseases require longer treatment than other acute diseases. Considering low- and middle-income countries the costs related with treatment and healing of chronic diseases may be more expensive and increase the risk of economic loss and impoverishment for families dealing with it. Treating chronic diseases is therefore expensive [2]. For example, in India the medical costs for treating diabetes for patients who visit private healthcare centers are 15% to 25% of their income. In comparison, in China, the cost for treating cancer in a hospital exceeds the annual income per-person [2]. Furthermore, considering developing countries, the burden is heavier as the costs for treating chronic diseases are handled by patients on their own.

It is evident that the management of healthcare is a significant socio-economic challenge for high-, low- and middle-income countries and it should be very organized and well managed so as to cope with the increasing healthcare costs [12].

The following sub-sections will elaborate briefly the challenges, concerns and issues that chronic diseases impose on the economical aspect of a developing country. Moreover, possible cost-effective solutions, such as the application of sensor based platforms, are also explored.

## 2.1 Economic Consequences of Chronic Diseases

The economic consequences of chronic diseases are evident not only considering the treatment costs, but also the high percentage of deaths caused by them, and the increased number of individuals that are diagnosed with chronic conditions. Related factors that can make the burden heavier are: the ageing population in developing regions, important changes in lifestyle, rapid urbanization rates, exposure to hypertension, less physical activities, etc.

Chronic diseases have an impact on labour-market performance, human-capital accumulation, saving decisions and consumption. Therefore, with the increasing evidence of the high number of people having chronic conditions, policymakers need to provide solutions to the growing risk of the disease in order to promote economic development. Moreover, considering that medical technologies and equipment for treatment of chronic diseases are becoming more advanced and sophisticated, treatment becomes more expensive [2]. This implies a financial burden, especially for developing countries to cope with the increasing medical cost and disease expansion.

There are also related costs that have an important impact on economic development, especially for low- and middle-income countries, and these are associated with the costs of obtaining treatment, transport to a medical center, or paying for medication and therapy. Therefore, the management and strategic analysis of the costs related with treatment of chronic diseases is beneficial for economic development of countries [13].

## 2.2 Cost-effective Solutions to Prevent Chronic Diseases

The prevention, management and curation of chronic diseases is believed to account higher percentages of the total spending of a nation's healthcare. A study from the World Health Organization (WHO) [3] concluded that developing countries should present a set of strategies to prevent, diagnose and treat chronic diseases, such as diabetes and heart disease, and other related diseases, for a cost of just 1.20$ per person per year.

In order to measure and provide a cost-effective solution while dealing with chronic diseases, one should consider all the actual costs, have reliable information related to health impacts on patients, and discount the future health expenses. Low-cost interventions that are required to facilitate disease prevention and mortality reduction were proposed by Russell [14], such as: increasing taxation on products that cause chronic conditions such as tobacco and alcohol, creating smoking-free areas, providing warnings and health information, launching public campaigns about healthy diets and physical activity. These methods, based on a study for 38 developing countries conducted by WHO [3], were successful in lowering the number of patients diagnosed with chronic diseases and the mortality rates of chronic ill individuals. However, recently, researchers have been intensively investigating the application of digital technologies, especially sensor based networks, in the health sector.

Digital technologies can improve the quality of health provision and reduce the corresponding expenses. For example, a Chronic Disease Self-Management Program (CDSMP) was presented [14] and a community-based patient self-management education course was described. This program enabled patients to reduce their visits to a medical center. In addition, the reduction of the number of hospitalized patients compared to the number of nights of hospitalization was significant ($p<0.05$). Another cross-sectional study made in 41 hospitals in Texas, concluded that hospitals that had implemented and applied advanced health information technology, had less concerns and complications, and reduced costs compared to others [15].

## 2.3 Living Cost and Related Health Expenses in Kosovo

Kosovo presents a small Balkan, developing country, whose health system is struggling to provide better services with very few resources and limited staff. The financial burden is one of the most important challenges that the health management is facing. Hence the funds distributed from the government to the Ministry of Health are limited compared to the needs of the system in general. From the total budget of the Ministry, 51% is spent in hospitals, 26% in municipalities, and 22% is used for other expenses [16]. It is worth to mention that only the half of total expenses in the health system is covered from governmental funds; the other half is financed from the patients [16]. The patient is required to pay for health services at the moment of requesting them. The average price for these services for a patient coming from rural areas is 50€ to 200€ including travelling costs [17]. On the other hand, the average wage in Kosovo is around 300€, the basic utilities expenses (electricity, heating, water, and garbage) are approximately 122€, expenses related to food are around 150€ per person, and transportation costs 20-30€ [18]. It is obvious that the sum of these expenses exceeds the overall income of a working individual. Taking into account that the majority of the population in Kosovo is unemployed, this situation worsens. Therefore, a huge difference is evident when comparing the health expenses with the available amount of money for Kosovo residents.

While comparing the total amount that a patient from rural areas pays for chronic diseases treatment in just one visit to the nearest medical center, with the total cost of a monitoring sensor-based system, it is shown that the visit costs more. For instance we consider two commercially available products, as introduced in the corresponding companies' websites. Zephyr is a wireless heart rate sensor compatible for Android phones which is used to measure: heart rate, RR interval (R wave to R wave interval), speed and distance. All you need is the Bluetooth communication to track the measurements. The cost for this application is around 65€ [19].

Another example is the ZIO patch, which consists of 2x5-inch patch which is able to record continuously patient's heart data up to 14 days. However, the ZIO patch presents an independent platform (no need to have smart phones as in the case of Zephyr), therefore it costs around 440€ [20]. Comparing the latter price with the approximate expenses of just one medical visit, it is evident that the platform price amounts to almost twice of the visit price. However, we should take into account that a chronic disease patient will have multiple visits to healthcare centers.

Using sensor based applications does not imply that there is no need for visiting medical centers, because the contact and advices from health professionals are essential. However, a patient may monitor his health parameters at home, and send them wirelessly, or save them in their phones and get medical advice if the values are out of range limits. This will lower the number of visits to medical centers, assist the patient in monitoring chronic parameters, react when these parameters do not indicate stable health conditions and eventually lower the treatment costs, since the sensor-based device is bought once, while almost the same cost is paid for each visit.

Having a better understanding of the possible impact of sensor-based technologies in lowering the cost for treatment and management of chronic diseases in a developing country, we consider important to research the user adoption with digital technology. The user, either medical staff or the patient itself, is a key factor of successful implementation of sensor-based applications. Therefore, a feasibility study of these two key factors can provide beneficial results for future works towards proposing a sensor-based architecture that will address the needs of patients with chronic disease who live in a developing country. Hence, in the next section we provide a brief literature research on the user perception and adoption with digital technology.

## 3. User Perception of Digital Technologies in Healthcare

Rapid developments and the widely-expanded use of digital technologies enabled the introduction of new applications and devices that provide opportunities for improving health services. Moreover, the cost reduction (based on the cost analysis), and proposal of affordable and more feasible solutions for healthcare services, impose the increased use of digital technology services in health system. Based on these observations, we consider as the next logical step to evaluate user opinion and perception. In this context, potential users are considered to be patients and the public, as well as healthcare professionals. The patients will use the sensor-based application for monitoring their chronic conditions, while the medical staff while need to have access to these readings and provide feedback. Therefore, e-health products must comply with a number of requirements including application accuracy, data security and integrity, patient privacy and control over data. In this context, there is a trade-off between the advantages and benefits that continuous monitoring tools offer and their functionality in order to meet not only the user's needs, but also meet their concerns or factors of usability.

A study by the U.S. Department of Health and Human Services in 2006 [21] found that, although many e-health tools were available at that time and were being adopted by users through the Internet, there were still questions and doubts about the utilization of these tools. Furthermore this study found that the consumer's most serious concerns were: data privacy, their attitude to use information technology, and their satisfaction with already developed e-health tools. Therefore, the study proposed that, considering the benefits that wireless monitoring tools offer, their usage should be supported to fulfil consumer-centric development, analysis and design, and concatenated them with public-private data exchange. This conclusion was supported also by [22] which suggested that having an understanding of the users' perspective may lead to successful implementation of e-health projects and innovations. More specifically, the study found that physicians and health managers primarily disagreed with the importance of privacy and security while applying Electronic Health Records (EHR). Regarding the importance of human resources, the most concerned groups were health professionals. These results showed that there exists a level of agreement and disagreement for specific factors within different user groups. These different viewpoints include the users' unique professional perceptions and objectives during the EHR process, therefore a challenge for successful implementation of the EHR process is evident, and this needs to be tackled.

A model for Health Information Exchange (HIE) for addressing population health needs was proposed in [23]. From the survey made in this study, the gathered results showed a positive percentage of users' attitude to use such systems.

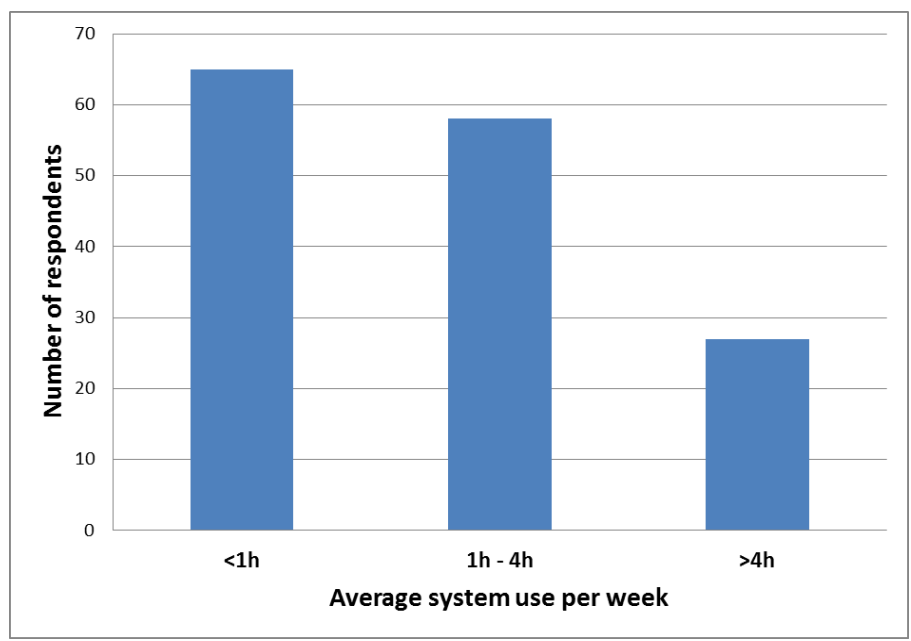

**Figure 2** Health Information Exchange system usage per week [23]

Figure 2 shows the users' responses (from a total number of 150 respondents) for using the health information exchange system within a weekly period. As the final bar of the figure shows, some respondents reported the use of the system for more than 4 hours. This study found that these results did not depend greatly on the users' age, for which the mean age was 43.8 years old (range 23-76).

We are aiming to investigate each user group's perception and possible factors that will impact the successful implementation of the e-health applications, and for that reason we have separated our work into two different points of views; healthcare providers and patients.

### 3.1 Healthcare Professionals Acceptance of E-health Services

The perception of clinicians towards e-health applications is important for the implementation and successfulness of such tools. In this context, the study presented in [24], briefly described the patient's requirements towards medical staff to use digital technologies. There were also studies investigating the health providers' attitude to using health information technologies.

A systematic literature review from various papers found in well-known databases such as PubMed, Medline, Cinahl, PsycInfo, ERIC, ProQuest Science Journals and EMBASE is presented in [25]. The researchers were focused only on papers which describe the healthcare providers' perception about the e-health services. Hence, this group of users (e.g. medical doctors) have a very important role in adopting e-healthcare services. So, even if patients would use e-healthcare tools, the medical professionals need to have access, read and understand the related data in order to provide a diagnosis for the individual patient.

In general, the process of adopting e-health services does not include only the communication infrastructure, but it requires from the corresponding staff to change their working style and adopt the new trends of health providing applications. The study presented in [26] found that the system's usefulness or the benefits that the e-health applications might provide, and the ease of use were the key factors while investigating the users' perception towards e-health tools. In contrast, the most constraining factors, similar to other studies, include: technical concerns, users' familiarity with digital technologies, application design etc. The study reviewed the health professionals' perception of some types of information and communication technologies, such as: electronic medical or health records; technology of retrieving medical information; hospital, clinical and nursing information systems; telemedicine; use of handheld devices; etc. Gagnon et al. [27] concluded that the related factors that would facilitate the application of e-healthcare had to do with the adoption of characteristics of health applications; while barriers were associated with the individual of professional perceptions of the users.

An electronic questionnaire was used in [28] to examine the physicians' perception toward health information technology by examining four models: the Technology Acceptance Model (TAM), the extended TAM, a psychosocial model, and an integrated model. The conducted survey had 157 participants and the results showed that 44-55% of physicians intended to use the electronic health records. However, the study suggested that the age, gender, the physicians' knowledge of using IT and previous experience should be taken into account. Similarly, another study [29] concluded that the use of primary health information systems would be fully accepted by nurses if they were believed to be trustworthy.

### 3.2 Patient Perception of Using E-healthcare Services

Various studies have found that patients have very similar concerns or attitudes towards digital technology as health professionals. The study in [24] investigated the email exchange between patients and clinicians. Patients considered that communication through email provide a useful, easy and fast way of consulting doctors, thus avoiding face-to-face consultations. This study underlined the facilities that email communication provide, such as: medical prescriptions, test reports, booking appointment or cancelling them, charts of chronic conditions, or informing patients for flu vaccinations or flu periods of year, etc. Similar studies, for example [30] and [31], noted that patients would like email access to their doctors. Moreover, research presented in [32] concluded that 90% of the participants support the possibility of using email for interacting with doctors, 56% indicated that this kind of communication influences the doctor's choice, while 37% of respondents were willing to use email exchange with their doctors, and they were willing to pay for it.

Possible factors that affect the acceptance of e-healthcare tools include: familiarity and ability to work with e-healthcare applications; patient interaction; motivation to use the specific tool based on usefulness of the application (this may be a factor which distinguishes the perception of e-healthcare for health professionals and patients in this case), data privacy and security, cost, ease of use, et al [33]. Another study [34] referred to the willingness to support and use new information technologies as Personal Innovativeness in Information Technology (PIIT).

Arising from the increasing aging population and the wide use of mobile health services (for example: continuous monitoring, surveillance, health related surveys, monitoring of patient and their records [35]) researchers in study [36] evaluated a research model about middle-age and older population perception about mobile healthcare. The survey results revealed that the higher their willingness to use e-healthcare services, the more intensive was their intention to practise them, for both middle-aged and older users. Another interesting conclusion has to do with the older users' acceptance of technology while avoiding traditional means of health services. The survey showed that the author's hypothesis that older users would prefer the continuity of traditional health services was contradicted by the results. The possible factors that may have led to this change may be the long waiting time for doctor consulting and visits [36]. However, there is still the technology anxiety and the ongoing introduction of new and sophisticated technological trends that patients are concerned about.

In order to investigate the heart failure patients and their readiness of using e-health applications for asserting their specific needs, a survey was conducted in [37]. The researchers examined the users' attitudes to use e-health program for telemonitoring and web based learning platforms. The results showed that patients with heart failure tended to use management e-health care programs.

In general, we can conclude that the reviewed literature indicates a convergence of several studies towards very similar results. The patients appear positive towards using information technology to monitor their living parameters and thus lower the number of visits to medical centers and enjoy improved health conditions. However, there are insecurities and concerns that should be taken into consideration.

To the best of our knowledge, most of the reviewed studies do not provide a concrete investigation on the user adoption of sensor-based technologies. Most of them provide user perception on the digital technology in general. Therefore, there is seen a need to analyse and investigate user opinions about the application of sensor technology in the health sector. It is very important to understand user concerns and requirements before proposing a sensor-based architecture.

## 4. Discussion

Patient and medical staff inclination to practise digital technology is indicated since it has the potential to help patients, decrease the number of visits to the medical centers, and lower the treatment costs [4]. However, the acceptance of sensor-based platforms in developing countries, as well as patients' and physicians' perception should be carefully considered. Adding to this the cost factor, the decision-making will be even more complicated.

Cost analysis showed that the overall living expenses in a developing country, such as Kosovo exceed the very minimum treatment costs that a chronic diseases patient needs. In addition, the lack of basic health insurance makes the burden heavier for chronic ill population. While considering developing countries, the provision of health insurance is considered quite important. However, there are a lot of developing countries that do not provide institutionally health insurance, in the context that employees or even citizens are not interested to have health insurance or cannot afford it, and employers are not obligated to provide it. In such cases, people can make individual health insurances, where the standard one (the cheapest) costs around 24€ per month (Kosovo case). This offer usually includes the most basic health services.

According to our literature review, there are just a few studies that have investigated the user perception of sensor-based platforms for monitoring chronic diseases in developing counties. Therefore, we intend to conduct an investigation towards users' perception of application of sensor based platforms in healthcare for the case of Kosovo. At this stage, we have defined our methods of research and formulated a structure of questions that we plan to utilize for our investigation. The idea of using both statistical methods, quantitative and qualitative one, relies on the following:

- Qualitative

  An important consideration should be provided to categorize the patients, in the context of their age, living area, and disease heaviness. In this way, understanding the average patients' economic situation will help possible interested parties to propose a system architecture and compare the existing expenses with the expected ones deriving from the usage of a sensor-based system. Moreover, it is useful to conduct interviews in order to examine users' attitude and willingness to accept such applications, so that their concerns can be addressed and possible requirements that are unique for developing countries can be revealed.

- Quantitative

  Questionnaires will be used to analyze the number of users that have already experience with sensor-based platforms for continuous monitoring of their health conditions. Corresponding results will help us make a correlation and provide specific statistics about the investigated patients. On the other hand, there may be users that have not used such applications; therefore we will 'measure' the perceived usefulness and the intention to learn how to use them.

The structure of questions planned for future investigation is based on user's categories, including patient age, gender, economic situation and living area; user experience with digital technology or sensor technology; and their perception and opinions of application of sensor-based technologies for monitoring chronic conditions. This structure of question applied through the qualitative or quantitative forms will enable the understanding of problems that a developing country is facing.

## 5. Conclusions

Digital technology presents a potential solution for continuous monitoring of patients with chronic diseases in low- and middle-income countries. Consequently, related research about the benefits and impact of investing in such technology is necessary. In this context, we have made a feasibility study on two of the most important factors that affect the successful implementation of sensor-based platforms; cost analysis and user perception.

Considering developing regions, cost is a key factor for effective application of sensor-based platforms. A comparison between literature-based data and the current economic, social and health situation in Kosovo, concluded that the overall management and traditional treatment cost of chronic conditions, exceeds the average income of working population in Kosovo. The application of sensor-based platforms is considered as a feasible and affordable method of dealing with chronic diseases. Taking into account the cost analysis outcome, we examined the user perception of e-health related digital technology. Based on a thorough literature review from various well-known databases, we conclude that there are few studies that investigate the health professionals' and chronic ill patients' perception of using sensor-based applications for continuous monitoring.

The conclusions presented in this paper, will lead our future research through investigating possible users of sensor-based applications on their attitude and willingness to use such applications. Therefore, we aim to identify possible problems and gaps that the application of sensor-based platforms may encounter in a developing country and treat them carefully in order to propose a feasible and suitable sensor-based architecture for chronic ill patients living there.

## Acknowledgments

The work presented in this paper is part of first author's PhD research at the South East European Research Centre (SEERC), overseas research centre of the University of Sheffield. The PhD studies were possible due to the scholarship provided by the Ministry of Education, Science and Technology of the Republic of Kosovo; and South East European Research Centre in Thessaloniki.

## Abbreviation list

CDSMP- Chronic Disease Self-Management Program
GDP- Gross Domestic Product
HER- Electronic Health Records
HIE- Health Information Exchange
ICT- Information and Communication Technology
PIIT- Personal Innovativeness in Information Technology
RR interval- R wave to R wave interval
SEERC- South East European Research Centre
TAM- Technology Acceptance Model
WHO- World Health Organization